\documentclass[prb,twocolumn,showpacs,preprintnumbers,amsmath,amssymb]{revtex4}

\usepackage{graphicx}
\usepackage{dcolumn}
\usepackage{bm}

\begin{document}

\title{Impurity-induced polar Kerr effect in a chiral $p$-wave superconductor}

\author{Jun Goryo}
 \email{jungoryo@s.phys.nagoya-u.ac.jp}
\affiliation{%
Department of Physics, Nagoya University, Furo-cho, Nagoya, 464-8602, Japan
}%

\date{\today}

\begin{abstract}
We discuss the polar Kerr effect (PKE) in a chiral $p$-wave ($p_x+i p_y$-wave) superconductor. It is found that the off-diagonal component of a current-current correlation function is induced by impurity scattering in the chiral $p$-wave condensate, and a nonzero Hall conductivity is obtained using the Kubo formula. We estimate the Kerr rotation angle by using this impurity-induced Hall conductivity and compare it with experimental results [Jing Xia et al., Phys. Rev. Lett. {\bf 97}, 167002 (2006)]. \end{abstract}


\maketitle



Recently, the quasi-two-dimensional (quasi-2D) superconductor Sr$_2$RuO$_4$ with $T_c=1.5$ K has attracted considerable attention  and it has been investigated extensively.\cite{Mackenzie-Maeno} It is plausible that the order parameter in Sr$_2$RuO$_4$ has the spin-triplet $p_x \pm i p_y$-wave symmetry.  One of the fascinating properties of this state is the spontaneous breaking of parity in a 2D sense ($p_x \rightarrow p_x$, $p_y \rightarrow - p_y$) and time-reversal symmetry due to the presence of nonzero chirality characterized by $l_z=\pm1$, where $l_z$ is the $z$-component of the relative orbital angular momentum of the Cooper pair. 

The polar Kerr effect (PKE), in which the direction of polarization of reflected linearly polarized light is rotated, has been known as an effective tool for understanding ferromagnetism.\cite{LRO} Because of the analogy between ferromagnetic order, for instance, with $s_z=1$ and chiral pair condensation with $l_z=1$, it is naively expected that the PKE is induced in the chiral $p$-wave state 
at zero field. In fact, the PKE has been observed in the superconducting state of Sr$_2$RuO$_4$.\cite{Xia} Up-to-date 
theoretical reports on the PKE in the chiral $p$-wave state are given in Ref.\cite{Lutchyn-Nagornykh-Yakovenko, Roy-Kallin}, in which interesting mechanisms have been proposed by the field theoretical approach; however, obtained results of the Kerr rotation angle are considerably smaller than the experimental results.  


Therefore, it is crucial to elucidate the fundamental nature of the PKE in the chiral $p$-wave superconductor.
We will show that a Kerr rotation angle comparable to that obtained experimentally is obtained by taking into account nonmagnetic and short-ranged impurity scattering of quasiparticles in a chiral $p$-wave condensate. It is also found that this impurity-induced PKE is suppressed or zero for any superconducting state other than the chiral $p$-wave state.\cite{Footnote} This result is contrary to the naive analogy with a ferromagnet, since the effect is not proportional to chirality analogous to magnetization, but suppressed in higher-chirality states with $l_z=\pm2,\pm3,\cdot\cdot\cdot$. The natural unit $\hbar=c=k_B=1$ is used throughout this study. 

We review the phenomenology of the PKE in time-reversal-symmetry-breaking superconducting systems, which is an extension of the argument for itinerant ferromagnetic systems.\cite{LRO,Aygyres} We also refer to the discussion of anyon superconductivity.\cite{Wen-Zee}  Suppose that $z>0$ is empty and $z<0$ is filled by the superconductor and incident light is linearly polarized and propagating along the $z$-direction perpendicular to the superconducting plane with a wavevector ${\bf q}=-\hat{\bf z} q_z$. The Maxwell equations inside the material are $\hat{\bf z} \nabla_z \times {\bf E}=-\partial {\bf B} / \partial t$ and $\hat{\bf z} \nabla_z \times {\bf B} ={\bf j} + \partial {\bf E} / \partial t$. It is obvious that for manifestly gauge invariant systems, current $j_i=\sum_{j=x,y}\sigma_{ij}E_j=-\sum_{j}\sigma_{ij}\left(\dot{A}_j+ \partial \phi/ \partial x_j\right)$, where $i=x,y$, $\sigma_{ij}$ is the conductivity tensor and $A_i$ and $\phi$ denote the vector and scalar potentials, respectively. In superconductors,  
\begin{eqnarray}
j_i=-\sum_{j} \left(\sigma_{ij}^{(v)} \dot{A}_j + \sigma_{ij}^{(s)} \frac{\partial \phi}{\partial x_j} \right),   
\label{current}
\end{eqnarray}
and in general, $\sigma_{ij}^{(v)}\neq\sigma_{ij}^{(s)}$ because of the spontaneous breaking of the gauge symmetry.\cite{Wen-Zee,footnote2} The scalar potential becomes redundant in this problem and we select a $\phi=0$ gauge. In this gauge, the Maxwell equation inside the superconductor is 
\begin{eqnarray}
\left\{\left(\omega^2 - q_z^2 + i \omega \sigma^{(v)}_{xx}(\omega)\right)\delta_{ij} + i \omega \epsilon_{ij} \sigma^{(v)}_{xy}(\omega)\right\}A_{j{\bf q}}(\omega)=0; 
\nonumber
\end{eqnarray}
here, we suppose that the long-wavelength limit ${\bf q}<<\xi^{-1}$, where $\xi$ is the coherence length of the superconducting order parameter and we omit the ${\bf q}$ dependence of the conductivity tensor. It is clear that there are two propagating modes
$
q_z^{\pm}=\sqrt{\omega^2+i\omega \sigma_{xx}^{(v)}(\omega)\pm \omega\sigma_{xy}^{(v)}(\omega)}.
$
In Ref.\cite{Wen-Zee}, the low-frequency limit $\omega<<2|\Delta|$ has been considered, and only the static values $\sigma_{ij}^{(v)}(\omega=0)$ have been considered, however, this value is not suitable for the experimental situation in the Ruthenate $\omega=0.8~eV>>2|\Delta| \simeq 10^{-4}~eV$.\cite{Xia} Following Ref.\cite{Aygyres}, we solve the Maxwell equation with an appropriate boundary condition at $z=0$ and obtain the Kerr rotation angle  
\begin{eqnarray}
\theta_K&=&-{\rm Im}\left(\frac{\omega (q_z^{+} - q_z^{-})}{\omega^2 - q_z^{+}q_z^{-}}\right), 
\label{Kerr}
\end{eqnarray}
which would be applicable in a wide frequency region. The factor $(q_z^{+} - q_z^{-})$ indicates that $\sigma_{xy}^{(v)}(\omega)$ is crucial to the PKE.

Let us calculate $\sigma_{ij}^{(v)}(\omega)$ with $\omega>>2| \Delta |$ in the chiral $p$-wave state. We use a quasiparticle Hamiltonian with a cylindrical Fermi surface that models the dominant $\gamma$-band in Sr$_2$RuO$_4$.\cite{Mackenzie-Maeno} The electromagnetic interaction and impurity scattering are also taken into account. In the Nambu representation $\Psi_{\bf p}=(c_{{\bf p} \uparrow},c^*_{-{\bf p} \downarrow})^{T}$, the Hamiltonian $H=H_0+H_{em}+H_{\rm i}$. The first part   
$H_0\equiv \int \frac{d^3p}{(2\pi)^3} \Psi^{\dagger}_{\bf p} {\bf g}_{\bf p} \cdot {\bm \tau} \Psi_{\bf p}$,
where ${\bf g}_{\bf p}=(Re \Delta_{\bf p}, -Im \Delta_{\bf p}, \epsilon_{\bf p}=\frac{{\bf p}^2}{2m_e} - \epsilon_F)$, ${\bm \tau}=(\tau_1,\tau_2,\tau_3)$ is the Pauli matrix in the Nambu space, and $\Delta_{\bf p}=|\Delta|(\hat{p}_x + i \hat{p}_y$) is the momentum-dependent part of a chiral $p$-wave gap function with $\hat{\bf p}={\bf p}/|{\bf p}|$. The second part 
$H_{em}=\int \frac{d^3q}{(2\pi)^3}  \left({\bf j}^{(p)}_{\bf q} + {\bf j}^{(d)}_{\bf q}\right) \cdot {\bf A}_{\bf q} $
is the electromagnetic interaction with the Fourier forms of the paramagnetic current 
${\bf j}^{(p)}_{\bf q}$$=-e\int \frac{d^3p}{(2\pi)^3}\Psi^{\dagger}_{\bf p}\frac{2 {\bf p} + {\bf q}}{2m_e}\Psi_{{\bf p}+{\bf q}}$
$\equiv $$-e \int \frac{d^3p}{(2\pi)^3} \Psi^{\dagger}_{\bf p}{\bm \gamma}_{{\bf p},{\bf p}+{\bf q}}\Psi_{{\bf p}+{\bf q}}$ and the diamagnetic current 
${\bf j}^{(d)}_{\bf q}$$=$$\frac{e^2}{2m_e}\int \frac{d^3p}{(2\pi)^3}\Psi^{\dagger}_{\bf p}\tau_3\Psi_{{\bf p}+{\bf q}} {\bf A}_{\bf q}$. 
The last part $H_{\rm i}=\nu_{\rm i} \int \frac{d^3q}{(2\pi)^3} \rho_{{\rm i} {\bf q}} \rho_{-{\bf q}}$  is 
the nonmagnetic impurity scattering with the Fourier components of the quasiparticle density $\rho_{\bf q} =\int \frac{d^3p}{(2\pi)^3} \Psi_{\bf p}^{\dagger} \tau_3 \Psi_{{\bf p}+{\bf q}}$ and impurity density $\rho_{{\rm i} {\bf q}}=\sum_j e^{i {\bf q}\cdot{\bf R}_j}$ (${\bf R}_j$: impurity site).  It is assumed that the scattering potential is short ranged and the $s$-wave channel is dominant. 


Let $\pi^R_{ij}(\omega)$ denote the Fourier component of the two-point retarded correlation of the paramagnetic current ${\bf j}_{{\bf q}\rightarrow 0}^{(p)}$. The Hall conducvitity is obained by the Kubo formula as\cite{Mahan} 
$\sigma_{xy}^{(v)}(\omega)=\pi^R_{xy}(\omega)/i \omega=\sum_{ij}\epsilon_{ij}\pi^R_{ij}(\omega)/2 i \omega$, where $\epsilon_{ij}$ is the totally antisymmetric tensor in the 2D plane. To obtain $\sum_{ij} \epsilon_{ij}\pi^R_{ij}(\omega)$, we calculate the Matsubara form $\sum_{ij} \epsilon_{ij}\pi_{ij}(i \omega_n)$, where $\omega_n=2 n \pi T$ is the bosonic Matsubara frequency at temperature $T$,  and use the relation $\sum_{ij} \epsilon_{ij}\pi_{ij}^R(\omega)=\sum_{ij} \epsilon_{ij} \pi_{ij}(i \omega_n \rightarrow \omega + i \gamma)$, where $\gamma$ denotes dissipation.\cite{Mahan} The Matsubara Green function for quasiparticles is given by  
\begin{eqnarray}
G_{\bf p}(i \epsilon_m)&=&\frac{1}{i \epsilon_m + {\bf g}_{\bf p}\cdot{\bm \tau}}
\equiv \left(
\begin{array}{cc}
{\cal{G}}_{\bf p}(i \epsilon_m)  & {\cal{F}}_{\bf p}(i \epsilon_m) \\
{\cal{F}}^*_{\bf p}(-i \epsilon_m)  & - {\cal{G}}_{\bf p}(-i \epsilon_m)        
\end{array}
\right), 
\nonumber
\end{eqnarray}
where $\epsilon_m=(2 m + 1)\pi T$. 

\begin{figure}
\includegraphics[width=\linewidth]{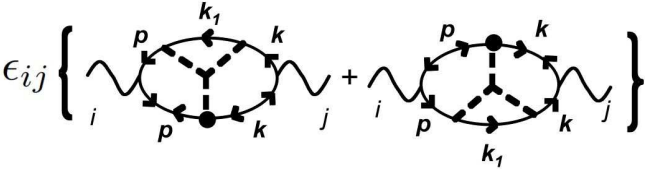}
\caption{\label{fig1} Leading diagrams of $\sum_{ij} \epsilon_{ij} \pi_{ij}(i \omega_n)$ . The lines with the arrows $->-$,  $-<->-$, and $->-<-$ denote quasiparticle Green functions ${\cal{G}}_{\bf p}(i\epsilon_m)$, ${\cal{F}}_{\bf p}(i\epsilon_m)$, and ${\cal{F}}^*_{\bf p}(i\epsilon_m)$, respectively, in $G_{\bf p}(i \epsilon_m)$. The lower lines have frequency $\epsilon_{m+n}=\epsilon_m + \omega_n$, while the upper lines have $\epsilon_m$. The dashed lines denote the impurity coupling $\nu_{\rm i}$. The scattering at the black dots yields the factor ${\rm Tr}(\tau_3 {\bf g}_{\bf p }\cdot {\bm \tau} {\bf g}_{\bf k} \cdot {\bm \tau})$ in Eq. (\ref{correlation}).}  
\end{figure}

In literature,\cite{Volovik,Goryo-Ishikawa,Furusaki,Read-Green,Horovitz}  it has been pointed out that there is no contribution of the one-loop diagram to the Matsubara form $\sum_{ij} \epsilon_{ij}\pi_{ij}(i \omega_n)$, i.e., the zeroth-order term of the impurity scattering   
since $\gamma_{i,{\bf p},{\bf p}}=p_i/m_e$.  
Then, vertex corrections must be taken into account to obtain nonzero contributions. 
The leading contribution is shown in Fig. 1, which is in the first order of the impurity concentration $n_{\rm i}$ and in the third order of the impurity strength $\nu_{\rm i}$. These diagrams are similar to the skew scattering diagrams in the extrinsic anomalous 
Hall effect.\cite{skew} We obtain   
\begin{widetext}
\begin{eqnarray}
&&\sum_{ij} \epsilon_{ij}\pi_{ij}(i\omega_n)
\nonumber\\
&=&n_{\rm i} \nu_{\rm i}^3 e^2\frac{T}{V^3}\sum_{m{\bf p}{\bf k}{\bf k}_1} \sum_{ij} \frac{\epsilon_{ij}}{2} 
\left\{{\rm Tr}\left(G_{\bf p}(i \epsilon_m)\gamma_{i,{\bf p},{\bf p}} G_{\bf p}(i \epsilon_{m+n}) \tau_3 G_{\bf k}(i \epsilon_{m +n}) \gamma_{j,{\bf k},{\bf k}}G_{\bf k}(i \epsilon_m) \tau_3 G_{{\bf k}_1}(i \epsilon_m) \tau_3\right)\right.
\nonumber\\
&&\left.+{\rm Tr}\left(G_{\bf p}(i \epsilon_m) \gamma_{i,{\bf p},{\bf p}}G_{\bf p}(i \epsilon_{m + n}) \tau_3 G_{{\bf k}_1}(i \epsilon_{m+n})\tau_3 G_{\bf k}(i \epsilon_{m+n})\gamma_{j,{\bf k},{\bf k}} G_{\bf k}(i \epsilon_m) \tau_3 \right)\right\}+{\cal{O}}(n_{\rm i} \nu_{\rm i}^4)
\nonumber\\
&=&
\frac{T}{V^3} \sum_{m{\bf p}{\bf k}{\bf k}_1} \frac{ n_{\rm i} \nu_{\rm i}^3 e^2 v_F^2 \hat{\bf p}\times\hat{\bf k} {\rm Tr}\left(\tau_3 {\bf g}_{\bf p} \cdot {\bm \tau} {\bf g}_{\bf k} \cdot {\bm \tau}\right) \omega_n (\epsilon_{m}+\epsilon_{m+n})^2(\epsilon_m \epsilon_{m+n}-E_{{\bf k}_1}^2)}{(\epsilon_m^2+E_{\bf p}^2)(\epsilon_{m+n}^2+E_{\bf p}^2)(\epsilon_{m}^2+E_{\bf k}^2)(\epsilon_{m+n}^2+E_{\bf k}^2)(\epsilon_m^2+E_{{\bf k}_1}^2)(\epsilon_{m+n}^2+E_{{\bf k}_1}^2)}
+{\cal{O}}(n_{\rm i} \nu_{\rm i}^4), 
\label{correlation}
\end{eqnarray}
\end{widetext}
where $E_{\bf p}=|{\bf g}_{\bf p}|=\sqrt{\epsilon_{\bf p}^2 + |\Delta_{\bf p}|^2}$ and $\epsilon_{m+n}\equiv \epsilon_m+\omega_n$. We use the cylindrical coordinate for representing momentum integration. As observed from the leading term of Eq. (\ref{correlation}), an azimuthal dependence $\hat{\bf p} \times \hat{\bf k}=\sin \theta _{pk}$ ($\theta_{pk}$: angle between ${\bf p}$ and ${\bf k}$) arises from the contraction of the vertex part $\sum_{ij} \epsilon_{ij} \gamma_{i,{\bf p},{\bf p}}\gamma_{j,{\bf k},{\bf k}}$, and  
\begin{eqnarray}
{\rm Tr}\left(\tau_3 {\bf g}_{\bf p} \cdot {\bm \tau} {\bf g}_{\bf k} \cdot {\bm \tau}\right)=-2 i \left(Re\Delta_{\bf p} Im\Delta_{\bf k} -  ({\bf p} \leftrightarrow {\bf k})\right)
\label{scattering}
\end{eqnarray}
arises from the impurity scattering at the black dots in Fig. \ref{fig1}.  In the chiral $p$-wave ($p_x+ip_y$-wave) state, ${\rm Tr}\left(\tau_3 {\bf g}_{\bf p} \cdot {\bm \tau} {\bf g}_{\bf k} \cdot {\bm \tau}\right) =2 i |\Delta|^2\sin \theta_{pk}$. Then,  the leading term of Eq. (\ref{correlation}) survives after the azimuthal integration. It is obvious that this term yields the second-order contribution of the gap amplitude $|\Delta(T)|^2\sim|\Delta(0)|^2(1-T/T_c)$ in the Ginzburg-Landau (GL) regime. By using GL expansion, we estimate the large $\omega_n$ part of the term to perform momentum integral first and Matsubara sum later.  By using analytic continuation and the Kubo formula, we obtain  
 \begin{eqnarray}
\sigma_{xy}^{(v)}(\omega)=\gamma_{BCS}^2\left(1-\frac{T}{T_c}\right)\frac{l_{\rm i}}{\xi_0}\left(\frac{\epsilon_{F}}{\pi \tau_0}\right)^{3/2}\frac{\sigma_{xy}^{(0)} }{(\omega + i / \tau_0)^3}, 
\label{Hall}
\end{eqnarray}
where $\sigma_{xy}^{(0)}=e^2/2\pi d$ ($d$: layer distance),  $l_{\rm i}=(n_{\rm i} d)^{-1/2}$, $\tau_0^{-1}=n_{\rm i} \nu_{\rm i}^2 N(0) /2$ ($N(0)=m/2 \pi d$: density of states at the Fermi surface per spin), $\xi_0=v_F/ \pi T_c $ is the superconducting coherence length at $T=0$, $\epsilon_{\rm F}$ is the Fermi energy, and $\gamma_{BCS}=|\Delta(0)|/T_c\simeq1.8$. 

In general, when we consider chiral states with $l_z=\pm1,\pm2,\cdot\cdot\cdot$, i.e., $(p_x+ip_y)^{l_z}$-wave states, Eq. (\ref{scattering}) becomes $2i|\Delta|^2\sin l_z \theta_{pk}$, and then, the leading term of Eq. (\ref{correlation}) vanishes after the azimuthal integration, except for $l_z=\pm1$, i.e., the chiral $p$-wave state. If we consider chiral states with a horizontal line node such as $(p_x+ip_y)^{l_z} p_z$, Eq. (\ref{scattering}) becomes $2 i |\Delta|^2 p_z k_z \sin l_z \theta_{pk} $ and the leading term also vanishes for any $l_z$ after integrating out the $z$-component of the momentums. For nonchiral ($l_z=0$) and time-reversal-breaking states, i.e., $d+is$-wave pairing, Eq. (\ref{scattering}) is proportional to $\cos 2\theta_p - \cos 2\theta_k$ and the leading term will vanish after the azimuthal integration. For nonchiral and time-reversal-symmetric states ($s$-wave, $d_{x^2-y^2}$-wave, ... ),  $\sigma_{xy}^{(v)}(\omega)$ should be zero because of the symmetrical reason. In fact, it is easy to confirm that Eq. (\ref{scattering}) becomes zero since the gap function can be made real for these states by using the $U(1)$ phase transformation of fermion fields. To sum up, in this impurity scattering mechanism, $\sigma_{xy}^{(v)}(\omega)$ is suppressed  or zero for any state other than the chiral $p$-wave state (see Footnote \cite{Footnote}).


The diagonal component $\sigma_{xx}^{(v)}(\omega)$ is effectively approximated by the Drude form
$\sigma_{xx}^{(v)}(\omega)\simeq\omega_p^2 \tau_0(1- i\omega \tau_0)^{-1}$
in the high-frequency limit $\omega>>2|\Delta|$\cite{Mattis-Bardeen}, where $\omega_p=\sqrt{n_e e^2/m_e}$ is the plasma frequency and $n_e$ is the electron number density.  In fact, this behavior of $\sigma_{xx}^{(v)}(\omega)$ has been verified experimentally in the superconducting state of Sr$_2$RuO$_4$.\cite{Katsufuji} 

Let us estimate $\theta_K$ from Eq. (\ref{Kerr}). We use suitable parameters for the experiment, i.e., $d=6.8~\AA$, $\xi_0=660~\AA$, $\omega=0.8~eV$, $\omega_p=1.3~eV$, $\epsilon_F=0.14~eV$, and $\tau_0^{-1}=6.6 \times 10^{-5}~eV$.\cite{Xia} The impurity-mean distance $l_{\rm i}$ is rather ambiguous, and we consider a variation $1000-5000~\AA$.\cite{Maeno} 
The estimated value of $\theta_K$ from Eq. (\ref{Kerr}) at $T/T_c=0.8$ is 6 nanorad for $l_{\rm i}=1000~\AA$ and 30 nanorad for $l_{\rm i}=5000~\AA$. The latter case agrees well with the measurement result of approximately 60 nanorad 
at the same temperature.\cite{Xia} 

\begin{figure}
\includegraphics[width=\linewidth]{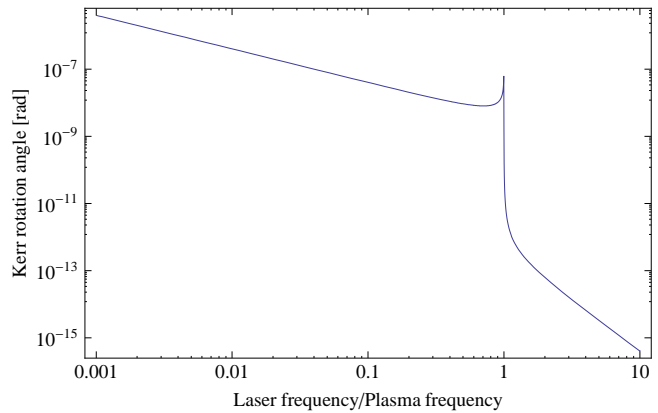}
\caption{\label{fig2} $\omega$ dependence of $\theta_K$ in Eq. (\ref{Kerr}) at $T/T_c=0.8$, obtained using parameters $d=6.8~\AA$, $l_{\rm i}=10000~\AA$, $\xi_0=660~\AA$,  $\omega_p=1.3~eV$, $\epsilon_F=0.14~eV$, and $\tau_0^{-1}=6.6 \times 10^{-5}~eV$.\cite{Xia} It is assumed that the cut off energy of the pairing interaction $\omega_D>10 \omega_p$. }  
\end{figure}

The frequency dependence of $\theta_K$ in Eq. (\ref{Kerr}) at $T/T_c=0.8$ is plotted in Fig. \ref{fig2}. 
We find that $\theta_K$ behaves as $\omega^{-1}$ for $\omega<\omega_p$, as indicated by Eqs. (\ref{Hall}) and (\ref{Kerr-a}). At $\omega=\omega_p$, $\theta_K$ shows a sharp peak and behaves as $\omega^{-3}$ for $\omega > \omega_p$. 

The temperature dependence of $\theta_K$ can be explained as follows. Eq. (\ref{Kerr}) is effectively approximated by 
\begin{eqnarray}
\theta_K\simeq \frac{\omega^2}{\omega_p^3} Re(\sigma_{xy}^{(v)}(\omega)), 
\label{Kerr-a}
\end{eqnarray}
when $\omega^2_p - \omega^2>>|\omega \sigma^{(v)}(\omega)|$ and $\omega \tau_0 >>1$ (these conditions are satisfied using suitable parameters in the experiment). The approximated form Eq. (\ref{Kerr-a}) indicates that $\theta_K$ depends linearly on the temperature in the GL region (see also Eq. (\ref{Hall})). This result is not in agreement with the experimental report;\cite{Xia} however, the error bars of the data are large and it appears to be difficult to discuss the temperature dependence. 

The obtained result might be overestimated since it is assumed that every scatterer has the same fixed potential strength rather than a distribution. We also suppose that the energy scale of the pairing interaction $\omega_D$ is well above the laser frequency  $\omega \simeq 0.8~eV$. Although the nature of the pairing interaction is not elucidated, $\omega_D$ is likely to be less than the used laser frequency. It is pointed out that the Hall conductivity would be suppressed for $\omega > \omega_D$.\cite{Roy-Kallin} However, we believe that the presented mechanism plays an important role in the PKE in Sr$_2$RuO$_4$.\cite{Xia}


We comment on previously proposed theories of the PKE in the chiral $p$-wave state. 
In Ref.\cite{Yakovenko}, $\sigma_{xy}^{(s)}$ in Eq. (\ref{current}) is obtained using the relation $\sigma_{xy}^{(s)}(\omega)=\frac{i}{2}\epsilon_{ij} \frac{\partial}{\partial q_i} \pi_{0j}^{R}(\omega,{\bf q})|_{{\bf q}=0}$, where $\pi_{0j}^{R}(\omega,{\bf q})$ is the correlation between charge density and current density. The calculation has been performed in the case without any impurity, and nonzero values of $\sigma_{xy}^{(s)}$ have been obtained; however, as shown in Eq. (\ref{Kerr}), $\sigma_{xy}^{(v)}(\omega)$ and not  $\sigma_{xy}^{(s)}(\omega)$ is responsible for the PKE. In Ref.\cite{Yakovenko}, it is implicitly assumed that $\sigma_{xy}^{(s)}(\omega)=\sigma_{xy}^{(v)}(\omega)$ in the high-frequency limit  (see the discussion below Eq. (17) in Ref.\cite{Yakovenko}. The same assumption has been made in Ref. \cite{Mineev}); however, this would not be true since $\sigma_{xy}^{(v)}(\omega)$ is zero without impurity scattering for an arbitrary frequency.\cite{Volovik,Goryo-Ishikawa,Read-Green,Furusaki}  This problem has also been pointed out in recent arguments.\cite{Lutchyn-Nagornykh-Yakovenko,Roy-Kallin} In Refs.\cite{Lutchyn-Nagornykh-Yakovenko,Roy-Kallin},   $\sigma_{xy}^{(s)}(\omega,{\bf q})$ has been discussed. A remarkable finding is that the finite ${\bf q}$ effect is responsible for the PKE (a similar argument has been provided in Ref.\cite{Goryo-IshikawaE-B}); however, the obtained Kerr angle is about  9 orders of magnitude smaller than that obtained experimentally.\cite{Lutchyn-Nagornykh-Yakovenko,Roy-Kallin} 




Let us turn to discuss the Chern-Simons term $\epsilon_{\mu\rho\nu}A_{\mu}\partial_{\rho}A_{\nu}$,  ($\mu,\rho,\nu=0,1,2$) which is generally induced in the low-energy and long-wavelength effective Lagrangian for gauge fields obtained by integrating out 2D electrons in a system with parity and time-reversal-symmetry breaking.\cite{DeserJackiwTempleton} The origin of the term is closely related to the parity anomaly.\cite{Redlich,Niemi-Semenoff,Ishikawa,Haldane} It has been pointed out that for the chiral $p$-wave  superconductor, a nonzero value of $\sigma_{xy}^{(s)}(\omega={\bf q}=0)$ is obtained without impurity scattering. At $T=0$, $\sigma_{xy}^{(s)}(0)=\frac{e^2}{4\pi}$;\cite{Volovik,Goryo-Ishikawa,Read-Green,Furusaki} the finite-temperature effect was investigated in the GL scheme\cite{Furusaki,chiralfeedback} (see also Ref.\cite{Horovitz}). This indicates that a part of the Chern-Simons term $\frac{\sigma_{xy}^{(s)}(0)}{2}\epsilon_{ij} \left(A_0 \partial_i A_j + A_i \partial_j A_0\right)$ is induced in the effective Lagrangian.\cite{Volovik,Goryo-Ishikawa} We can conclude that this part has an {\it "intrinsic"} origin since it is induced without impurity scattering, while $\sigma_{xy}^{(v)}(0)=0$, and the other part of the Chern-Simons term $\epsilon_{ij} A_i \partial_0 A_j$ is not induced.\cite{Volovik,Goryo-Ishikawa,Read-Green,Furusaki}
As we have emphasized, $\sigma_{xy}^{(v)}(\omega)$ becomes nonzero when we take into account vertex corrections from impurities,  shown in Fig. \ref{fig1}. From the leading term of Eq. (\ref{correlation}), we can observe that in the static limit, the second-order contributions of the gap amplitude vanish occasionally, and the forth-order contributions give 
$$
\sigma_{xy}^{(v)}(0)=\sigma_{xy}^{(0)}\frac{127\zeta(7)}{128 \pi^4} \frac{l_i}{\xi_0}\left(\frac{\epsilon_F}{\pi \tau_0 T_c^2}\right)^{\frac{3}{2}} \left(1-\frac{T}{T_c}\right)^2 \gamma_{BCS}^4
$$ 
near $T_c$. This indicates that the other part $\frac{\sigma_{xy}^{(v)}(0)}{2}\epsilon_{ij} A_i \partial_0 A_j$ has an {\it "extirisic"} origin. 

In summary, we have discussed the PKE in a chiral $p$-wave superconductor. We have estimated the contribution of impurity scattering in the chiral $p$-wave condensate. The skew-scattering-type diagrams (see Fig. \ref{fig1}) show the leading contribution. In this impurity-induced mechanism, the PKE would be suppressed or zero for any state other than the chiral $p$-wave state (the possibility of nonunitary pairing is not taken into account (see Footnote \cite{Footnote}). 


The author is grateful to D. S. Hirashima, H. Kontani, Y. Maeno, and V. M. Yakovenko for useful discussions and comments. This work was supported by Grant-in-Aid for Scientific Research (No. 19740241) from the Ministry of Education, Culture, Sports, Science and Technology.


\begin{thebibliography}{99}
\bibitem{Footnote} The PKE would occur in the nonunitary (spin-polarized) pairing state;\cite{Sigrist-Ueda}, however, this state is not taken into account here since it is difficult to imagine that such a state arises at zero field without some special condition, for example,  the superconducting phase is surrounded by a ferromagnetic phase. Sr$_2$RuO$_4$ does not have a ferromagnetic phase.
\end{thebibliography}
\end{document}